\documentclass[fleqn,usenatbib]{mnras}

\usepackage{newtxtext,newtxmath}

\usepackage[T1]{fontenc}
\usepackage{ae,aecompl}

\usepackage{graphicx}	
\usepackage{amsmath}	
\usepackage{amssymb}	
\usepackage{multirow}
\usepackage[normalem]{ulem}

\title[Li abundances in M67]{Li abundances for solar twins in the open cluster M67}

\author[M. Carlos et al.]{
Mar\'ilia Carlos,$^{1}$\thanks{E-mail: marilia.carlos@usp.br}
Jorge Mel\'endez,$^{1}$,
Jos\'e-Dias do Nascimento Jr.$^{2,3}$ and \newauthor Matthieu Castro$^{2}$
\\
$^{1}$Departamento de Astronomia, IAG, Universidade de S\~ao Paulo, Rua do Mat\~ao 1226 S\~ao Paulo, 05509-900, Brazil\\
$^{2}$Dep. de F\'isica, Universidade Federal do Rio Grande do Norte, CEP: 59072-970 Natal, RN, Brazil\\
$^{3}$Harvard-Smithsonian Center for Astrophysics, Cambridge, MA 02138, USA
}

\date{Accepted: 2019 December 10. Received: 2019 December 4; Received: 2019 October 15}

\pubyear{2019}

\begin{document}
\label{firstpage}
\pagerange{\pageref{firstpage}--\pageref{lastpage}}
\maketitle

\begin{abstract}

We determine lithium (Li) abundances for solar twins in the M67 open cluster to add valuable information about the correlation between Li depletion and stellar age and, then, better understand stellar structure and evolution. We use high resolution and good signal-to-noise ratio spectra to characterize Li depletion in three solar twins from M67, using spectral synthesis in the region of the asymmetric 6707.75 \AA \, Li {\sc I} feature. The mean Li abundance value of A(Li)$=1.6\pm0.2$ dex for our sample of M67 solar twins (our three stars plus a fourth solar twin from a previous analysis in the literature) presents Li abundance expected for its age. Also, the scatter estimated from the standard deviation of the Li abundances in this work is similar to the typical scatter found in a sample of field solar twins presented in the literature.

\end{abstract}




\section{Introduction}

Because of its fragile nature, Li is an excellent astrophysical tracer of transport mechanisms and mixing within and below the convective zone of solar-type stars. This element is destroyed at temperatures near  $2.5\times10^6$ K, when is transported to the inner regions of a star as a result of its convective motions and extramixing processes.

Studies of the correlation between Li depletion and stellar age can help to constrain non-standard stellar evolution models and, thus, improve our knowledge on the structure and evolution of the Sun and Sun-like stars. These non-standard stellar evolution models take into account different theoretical physics approaches to try to explain the low content of lithium in the Sun and Sun-like stars. Models explore different physics, such as gravity waves  \citep{charbonnel/talon/05}, rotation-induced mixing and diffusion \citep{donascimento/09}, overshooting and gravitational settling \citep{xiong/deng/09}, and rotation-driven turbulent diffusion \citep{denissenkov/10}.



Several works in the literature discuss lithium depletion with stellar age \citep{sestito/randich/05,ford/05,monroe/13,carlos/16,carlos/19}, and many of them study field stars in which the age determination may yield considerable uncertainties. Thus, it is imperative to analyse stars in clusters due to their better age estimates. 

The M67 (NGC 2682) open cluster presents an ample range of studies in the literature. Its age and metallicity make this cluster a good asset to assess the stellar evolution of the Sun and solar-like stars. The estimated age of M67 found in the literature varies from 3.4 to 5.4 Gyr \citep[e.g.,][]{vandenberg/04,sarajedini/09,magic/10,gaia_col/18}, while its metallicity varies within $-0.1\lesssim\rm{[Fe/H]}\lesssim0.1$ \citep[e.g.,][]{fan/96,onehag/11,gaia_col/18,liu/16,liu/19,souto/19II}.

Although the works of \cite{pasquini/08} and \cite{castro/11}  have measured Li abundances for several solar-like stars in M67, the spectral resolution of about 17000 (Fiber Large Array Multi Element Spectrograph -- FLAMES/GIRAFFE spectrograph; \citealt{pasquini/02}) is not enough to resolve the 6707.43 \AA \, Fe I and 6707.75 \AA \, Li I lines and, thus, compromise the final Li abundances achieved on their studies. \cite{onehag/11} and \cite{liu/16} analysed one solar twin  at high resolution ($\rm{R}\sim50000$), adding valuable information about Li abundances in clusters. As there is a lack of high-resolution spectra of M67 solar twins to study Li depletion, it is critical to acquire more high-quality data to the discussion.   


In this paper we discuss the Li depletion and its scatter in the M67 cluster, using high resolution and good signal-to-noise ratio (S/N) spectra for three solar twins.

\section{Sample and Analysis}


The sample consists of three solar twins  observed with the Gemini Remote Access to CFHT ESPaDOns Spectrograph\footnote{For more information see http://www.gemini.edu/sciops/future-instrumentation/graces/documents} \citep[GRACES,][]{chene/14} at the Gemini North Observatory with high resolution (R$\sim$45000) and moderately high S/N (S/N$\sim$100) on the night of 2016 December 23, 24, 25, and 27  (observation program GN-2016B-Q-36). The sample, presented in Table \ref{log_tab}, was chosen from a list built with M67 objects that have high membership probabilities, calculated using proper motion from \cite{yadav/08}, previous spectroscopic studies by \cite{pasquini/08}, and further studies of radial velocity by \cite{geller/15}. Furthermore, we verified objects that may have about $1 \mathrm{M}_{\odot}$ and some photometric modulation from Kepler K2, as analysed by \cite{barnes/16} and \cite{gonzalez/16}.

\begin{table*}
	\centering
	\caption{Coordinates, V magnitudes, radial velocities, proper motions and Epic names for our sample.}
	\label{log_tab}
	\begin{tabular}{lccccccr} 
		\hline
		\hline
		Star & $\alpha$(J2000) & $\delta$(J2000) & V$^\star$ (mag) & v$_{\rm{rad.}}^{\star}$ (km.s$^{-1}$) & P$_{V}$(\%)$^{\dagger}$ & P$_{\mu}$(\%)$^{\ddagger}$ &  Epic name\\
		\hline
		Cl$^\ast$ NGC 2682 FBC 1877 & 08 50 21.5753 & +11 50 23.1764 & 14.52 & 34.57 & 97 & 99 & 211413212\\
		Cl$^\ast$ NGC 2682 YBP 285  & 08 51 23.8697 & +11 38 52.1557 & 14.46 & 34.01 & 98 & 98 & 211400500\\
		Cl$^\ast$ NGC 2682 YBP 1303 & 08 50 56.6633 & +11 49 54.6772 & 14.64 & 37.27 & 98 & 99 & 211412674\\
		\hline
		\hline
	\end{tabular}
	\\
		{\bf Notes.} $^{\star}$Data from \cite{geller/15}. $^{\dagger}$Membership probability from \cite{geller/15}. \newline 
		$^{\ddagger}$\cite{yadav/08} proper motion membership probability.
\end{table*}

\begin{figure}
\begin{tabular}{c}
	\includegraphics[width=0.5\textwidth]{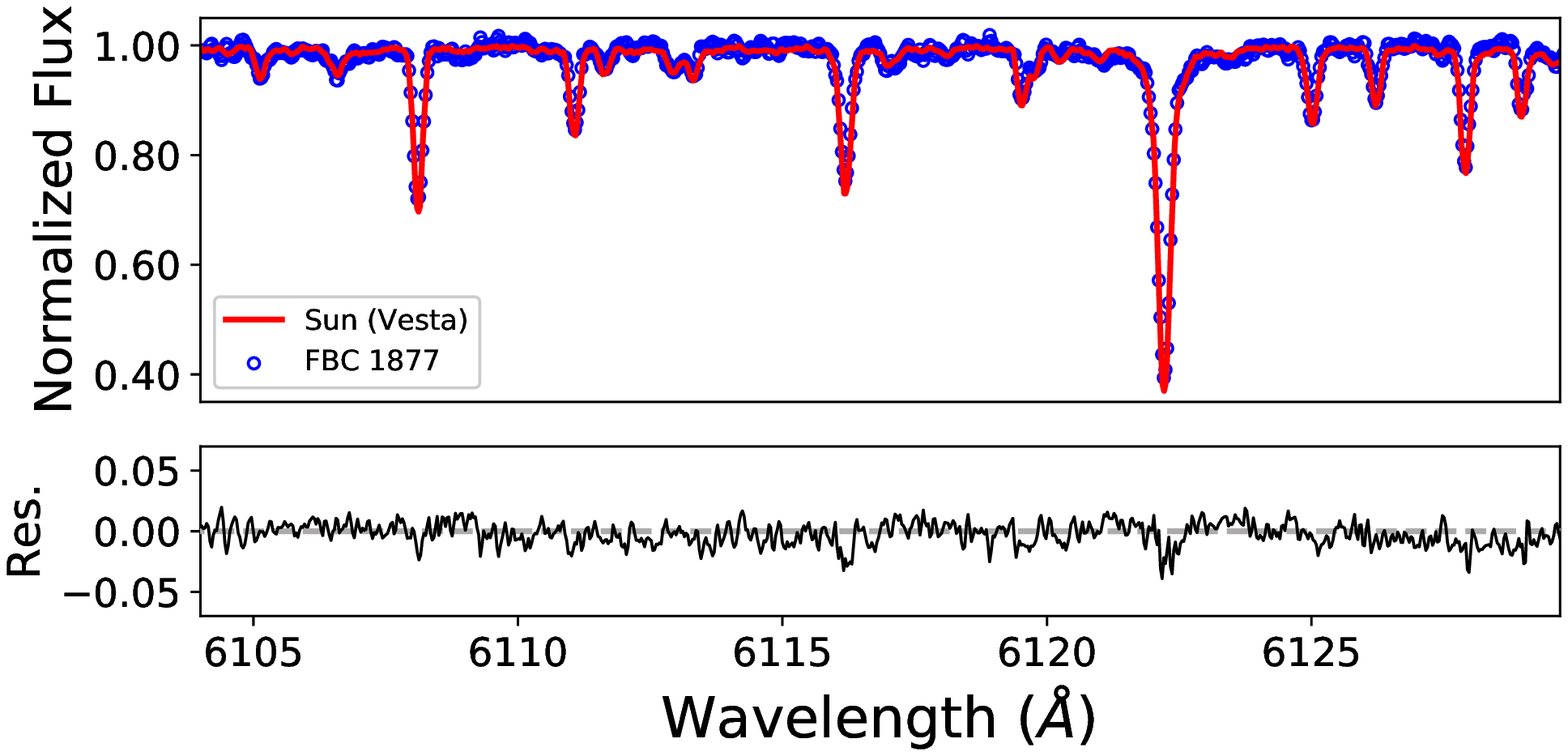} \\
	\includegraphics[width=0.5\textwidth]{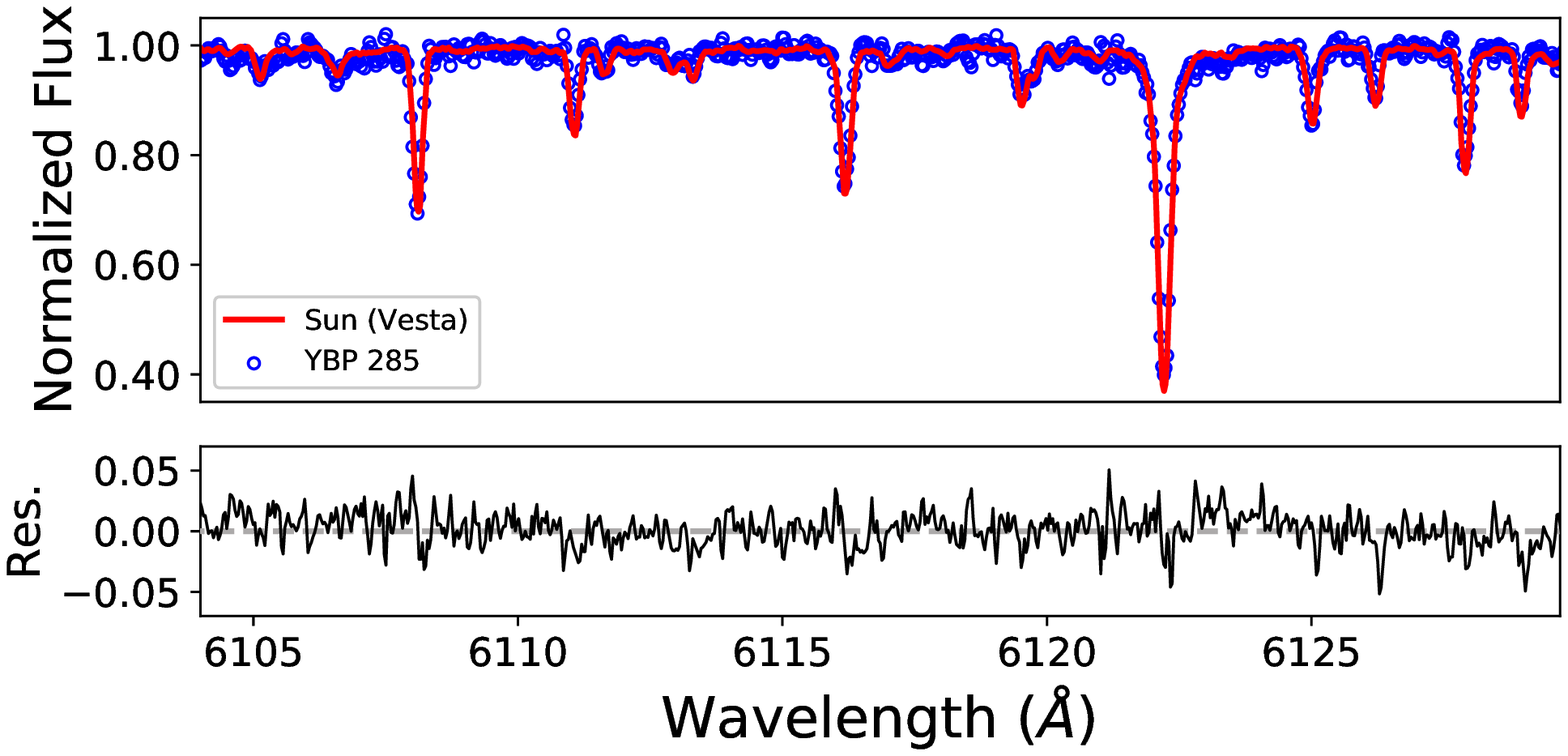} \\
	\includegraphics[width=0.5\textwidth]{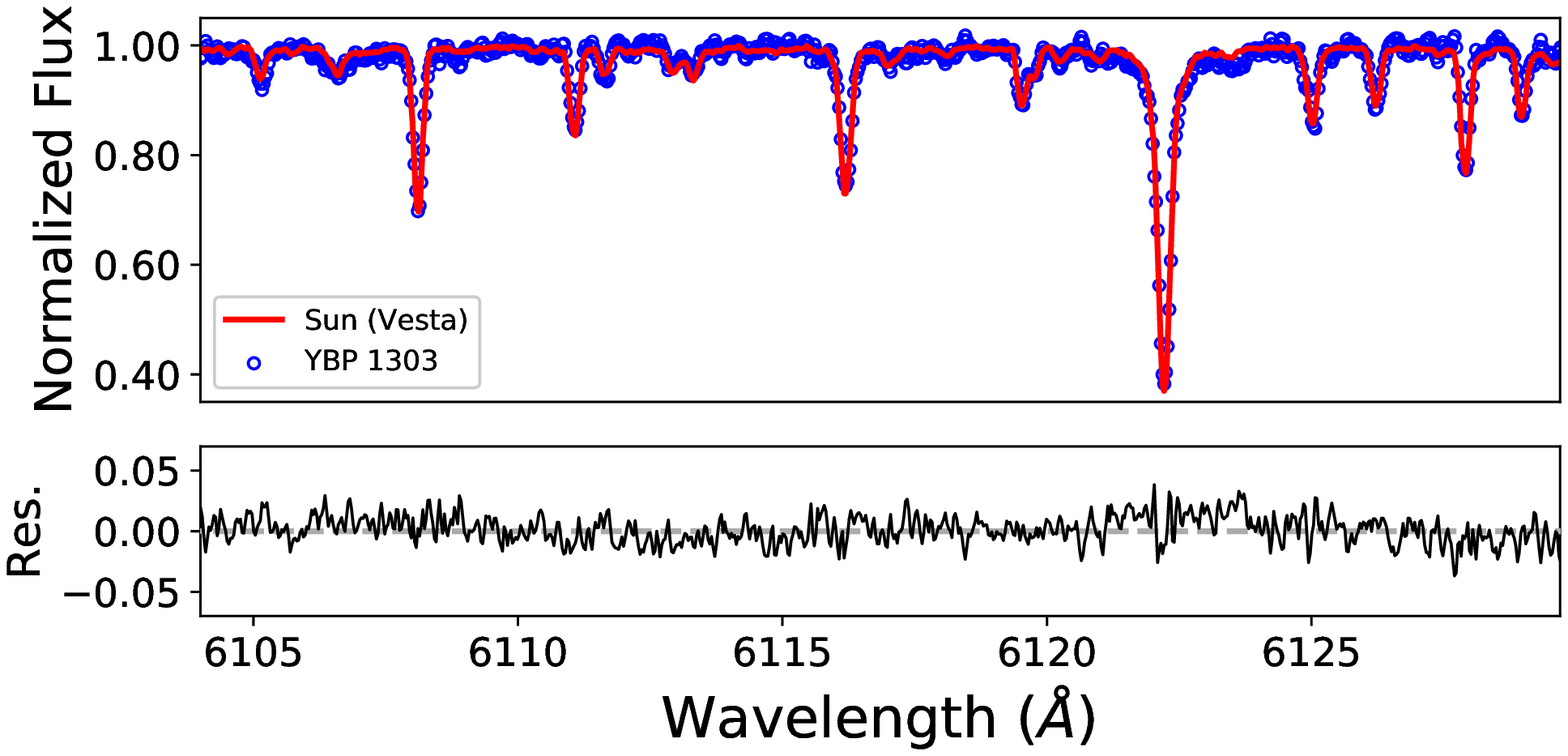} 
\end{tabular}    
    \caption{Solar spectrum (red line) in comparison with three solar twins from M67 (open blue circles). Their respective residuals are shown in black.}
    \label{solar_twins}
\end{figure}

The spectra were reduced and normalized using the OPERA pipeline \citep{martioli/12,teeple/14}. Fig. \ref{solar_twins} shows the comparison between the M67 solar twins candidates and the Sun around 6115 \AA. The solar spectrum used for comparison was taken by the GRACES spectrograph, and was observed in reflected sunlight from the asteroid Vesta. The residues, which indicate the difference between the M67 stars and the Sun, have a value of about $\sim$ 1\% and illustrate the similarity that is expected for solar twins.

The stellar parameters were calculated by using two different methods to verify their consistency. Through a differential spectroscopic analysis of equivalent widths using the {\sc IRAF}\footnote{http://iraf.noao.edu/} task {\sc SPLOT}, and also via photometry, with data from the literature. Both analysis were done with the q2 python package \citep{ramirez/14b}.

In the differential spectroscopy analysis, the effective temperature, surface gravity and [Fe/H] were estimated by using Fe {\sc I} and Fe {\sc II} lines in a differential line-by-line method to achieve excitation and ionization equilibrium balance \citep[e.g.,][]{melendez/14b,bedell/14}. Then, the masses were calculated by using the stellar parameters derived from the spectroscopic method using Yonsei-Yale isochrones \citep{yi/01,kim/02}. 

In order to verify our spectroscopic stellar parameters, the effective temperature, surface gravity and mass were also derived using photometric data from the literature. 

The $V, J, H, K_S$ magnitudes and the $B-V$ colour \citep{geller/15,pace/12,cutri/03} were used with the photometric calibration from \cite{casagrande/10} to estimate the effective temperatures (T$_{\rm{eff}}$). We adopted the M67 reddening $E(B-V)=0.037$ from the \cite{gaia_col/18}. The reddening ratios used to correct the other colours, $k=E(colour)/E(B-V)$, were taken from \cite{ramirez/05a}. 

The stellar parameters mass and surface gravity (log g) were also estimated by using the Yonsei-Yale isochrone set \citep{yi/01,kim/02}. We adopt the photometric T$_{\rm{eff}}$ previously mentioned, [Fe/H] from our spectroscopic analysis, the V  magnitude from \cite{geller/15}, the $E(B-V)$ reddening from the \cite{gaia_col/18}, and parallax from the Gaia Data Release 2 (DR2) archive with the zero point offset correction (we added 0.03 mas) suggested by \cite{lindegren/18/gaiadr2}.


The stellar parameters presented in Table \ref{all_par} taken both by photometry and spectroscopy show the consistency between the two different methods used in this work and are compatible within 1$\sigma$. Our parameters also confirm that the sample consists of only solar twins as the effective temperature is within T$_{\mathrm{eff},\odot}\pm100$ K, surface gravity within log g$_{\odot}\pm0.1$, and metallicity  within [Fe/H]$_{\odot}\pm0.1$.


\begin{table*}
\centering
\caption{Stellar parameters obtained through photometry and high-resolution spectroscopy.}
\label{all_par}
\begin{tabular}{l|ccrc|ccccr}
\hline \hline
\multirow{2}{*}{Star} & \multicolumn{3}{c|}{Photometry}                                                                                                                                                                                                                        & \multicolumn{6}{c|}{Spectroscopy}                                                                                                                                                                                                                                                                                                                                           \\ \cline{2-4} \cline{6-10} 
                      & \multicolumn{1}{c}{\begin{tabular}[c]{@{}c@{}}Mass \\ (M$_{\odot}$)\end{tabular}} & \multicolumn{1}{c}{\begin{tabular}[c]{@{}c@{}}T$_{\mathrm{eff}}$ \\ (K)\end{tabular}} & \multicolumn{1}{c|}{\begin{tabular}[c]{@{}c@{}}log g\\ (dex)\end{tabular}} & & \multicolumn{1}{c}{\begin{tabular}[c]{@{}c@{}}Mass \\ (M$_{\odot}$)\end{tabular}} & \multicolumn{1}{c}{\begin{tabular}[c]{@{}c@{}}T$_{\mathrm{eff}}$ \\ (K)\end{tabular}} & \multicolumn{1}{c}{\begin{tabular}[c]{@{}c@{}}log g \\ (dex)\end{tabular}} & \multicolumn{1}{c}{\begin{tabular}[c]{@{}c@{}}v$_t$\\ (km.s$^{-1}$)\end{tabular}} & \multicolumn{1}{r}{{[}Fe/H{]}} \\ \hline
Our work & & & & & & & &\\
FBC 1877 & 1.04$^{+0.03}_{-0.03}$ & 5843 $\pm$ 65 & 4.41$^{+0.04}_{-0.04}$ & & 1.04$^{+0.02}_{-0.02}$ & 5840 $\pm$ 24 & 4.46 $\pm$ 0.06 & 1.02 $\pm$ 0.05 & 0.06 $\pm$ 0.02 \\ 

YBP 285 & 1.05$^{+0.03}_{-0.04}$ & 5874 $\pm$ 105 & 4.42$^{+0.04}_{-0.05}$ & & 1.06$^{+0.03}_{-0.02}$ & 5883 $\pm$ 47 & 4.51 $\pm$ 0.08 & 1.15 $\pm$ 0.08 & 0.07 $\pm$ 0.03 \\

YBP 1303 & 0.99$^{+0.03}_{-0.02}$ & 5750 $\pm$ 39 & 4.41$^{+0.04}_{-0.04}$ & & 1.01$^{+0.02}_{-0.02}$ & 5768 $\pm$ 36 & 4.50 $\pm$ 0.06 & 1.14 $\pm$ 0.07 & 0.04 $\pm$ 0.03 \\
\hline
\cite{liu/16} & & & & & & & &\\
YBP 1194 & -- & -- & -- & & (1.01$^{+0.02}_{-0.02}$)$^{\dagger}$ & 5786 $\pm$ 13 & 4.46 $\pm$ 0.02 & 1.04 $\pm$ 0.02 & -0.005 $\pm$ 0.010\\
\hline \hline
\end{tabular}
{\bf Notes.}$^{\dagger}$ Our mass calculation  using \cite{liu/16} data.
\end{table*}

The stars FBC 1877 and YBP 285 have rotational periods registered in the literature. The work of \cite{barnes/16} found rotational periods of $24.4\pm2.5$ and $26.9\pm1.5$ d for FBC 1877 and YBP 285 respectively; within the errors these values are expected  for solar twins at this age \citep{lorenzo-oliveira/19}.


The Li abundances were derived with the same method applied in \cite{carlos/16,carlos/19}, using spectral synthesis of the asymmetric 6707.75 \AA \, Li {\sc I} blend with the aid of the July 2014 version of the 1D local thermodynamic equilibrium (LTE) code {\sc MOOG} \citep{sneden/73} and the Kurucz grid of {\sc ATLAS9} model atmospheres \citep{castelli/04}. The line list of the Li region from \cite{melendez/12}, which includes the components of the Li feature and blends from atomic and molecular (CN and C$_2$) lines, was adopted.

The stellar parameters  employed for the spectral synthesis are those from the spectroscopic analysis, due to the smaller errors in comparison to those from the photometric method. In order to compare the synthetic and observed spectra, it is important to take into consideration the line broadening effects due to macroturbulence (v$_{\rm{macro}}$) and projected rotational  velocities (v$\sin i$). We estimated the v$_{\rm{macro},\odot}=4.1\, \rm{km.s}^{-1}$, adopting  v$\sin i=1.9\, \rm{km.s}^{-1}$, for our GRACES solar spectrum.

For the three stars in the sample, the macroturbulence velocity was calculated by the Equation (1) of \cite{dos_santos/16}, and the projected rotational velocity was determined by analysing the line profile of the Fe I 6027.050 \AA, 6093.644 \AA, 6151.618 \AA, 6165.360 \AA, 6705.102 \AA, and Ni I 6767.772 \AA \, lines. 

\begin{figure}
	\includegraphics[width=\columnwidth]{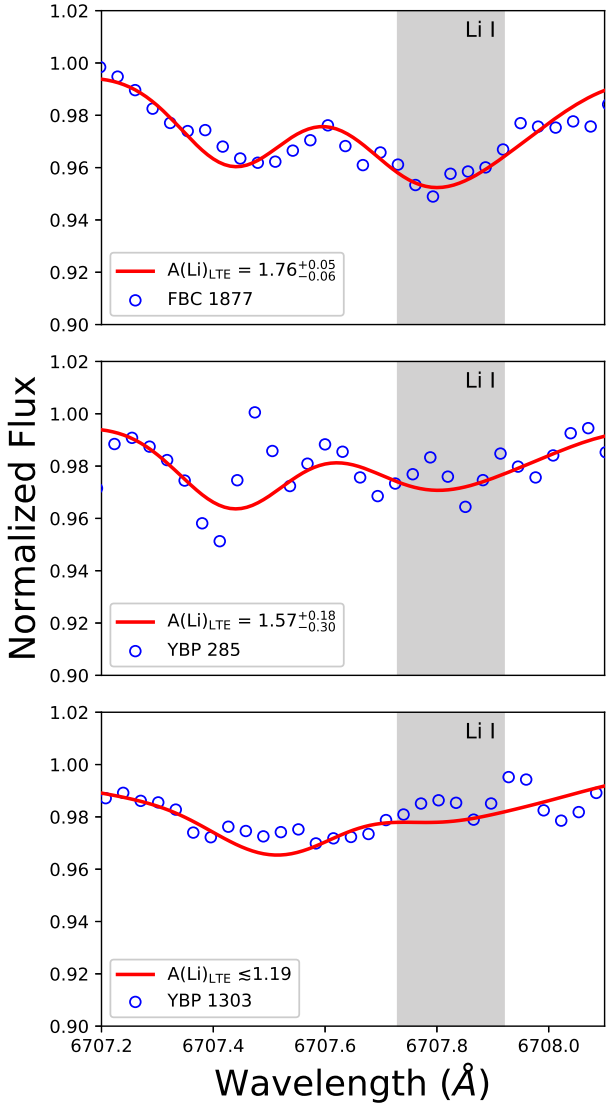}
    \caption{The red solid line represents the spectral synthesis in comparison with the observed data (open blue circles) for the M67 stars FBC 1877 (top panel), YBP 285 (middle panel), and YBP 1303 (bottom panel).}
    \label{all_stars_li}
\end{figure}

Fig. \ref{all_stars_li} shows the spectral synthesis of  6707.75 \AA \, Li I line region for the three stars of our sample. The LTE Li abundances were corrected to non-LTE (NLTE) abundances using the INSPECT data base\footnote{\url{www.inspect-stars.com} (version 1.0).}, based on NLTE calculations by \cite{lind/09}. The errors were estimated by considering the uncertainties in the stellar parameters, the rms deviation of the observed line profile relative to the synthetic spectra, and the continuum setting. The LTE and NLTE values of $\rm{A(Li)}=\log(\rm{N}_{Li}/\rm{N}_{H}) + 12$, with their respective errors, are presented in Table \ref{li_tab}.

\begin{table}
	\centering
	\caption{Lithium abundances of solar twins in M67.}
	\label{li_tab}
	\begin{tabular}{lcr} 
		\hline
		\hline
		Star & \multicolumn{1}{c}{\begin{tabular}[c]{@{}c@{}}A(Li) LTE\\(dex)\end{tabular}} & 
		\multicolumn{1}{c}{\begin{tabular}[c]{@{}c@{}}A(Li) NLTE\\(dex)\end{tabular}} \\
		\hline
		Our work, R$=45000$ & & \\
		FBC 1877 & 1.76$_{-0.06}^{+0.05}$ & 1.79$_{-0.06}^{+0.05}$\\
		YBP 285  & 1.57$_{-0.30}^{+0.18}$ & 1.60$_{-0.30}^{+0.18}$\\
		YBP 1303 & $\lesssim$ 1.19 & $\lesssim$ 1.24\\
		\hline
		\cite{liu/16}, R$=50000$ & & \\
		YBP 1194 & 1.32$^{+0.08}_{-0.07}$ & 1.36$^{+0.08}_{-0.07}$\\
		\hline
		\cite{castro/11}, R$=17000$ & & \\
		YBP 285 & 0.6 & -- \\
		YBP 1303 & 1.2 & -- \\
		\hline
		\cite{pasquini/08}, R$=17000$ & & \\
		YBP 285 & 0.6 & $<0.6$\\
		YBP 1303 & 0.5 & $<0.6$ \\
		\hline
		\hline
	\end{tabular}
\end{table}

\section{Discussion}

Two of the three stars in our sample have Li abundances in the literature. The works of \cite{pasquini/08} and \cite{castro/11} measured the Li abundances for the stars YBP 285 and YBP 1303 using spectra from the multi-object FLAMES/GIRAFFE spectrograph at the UT2/Kueyen ESO-Very Large Telescope \citep[VLT;][]{pasquini/02}, which have a resolution that do not separate the 6707.43 \AA \, Fe {\sc I} and 6707.75 \AA \, Li {\sc I} lines (R$\sim$17000). From Fig. \ref{all_stars_li} it is possible to notice that these two lines are resolved in our data (S/N is about the same, but our resolving power is about three times higher) and, thus, our study likely achieved better Li abundance estimates. Notice that for star YBP 285, our more precise result is much higher than in \cite{castro/11}, but for YBP 1303 we get about the same result; note that the \cite{castro/11} paper is a reanalysis of the FLAMES/VLT spectra by \cite{pasquini/08}, but using spectral synthesis instead of equivalent widths of the Li blend. For details on the wavelength position for each species around the 6707.75 \AA \, Li {\sc I} line region, see fig. 2 of \cite{carlos/16}. For the first time Li abundance was estimated for the star FBC 1877. 

\begin{figure}
	\includegraphics[width=\columnwidth]{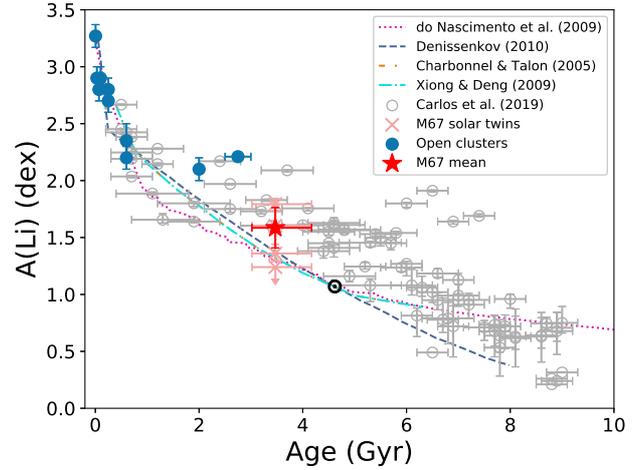}
    \caption{Li vs. age correlation. The grey open circles represent field solar twins analysed by \protect\cite{carlos/19}. Blue filled circles show open clusters studied in the literature (references are shown in the text). Single solar twins from M67 are displayed by light red crosses and its respective mean value is shown by the red star. The Sun is represented by its usual symbol.}
    \label{li_age}
\end{figure}

In addition, the work of \cite{liu/16} analysed the solar twin Cl$^{\ast}$ NGC 2682 YBP 1194  (spectroscopic stellar parameters are reproduced in Table \ref{all_par}), a star also studied by \cite{onehag/11}, using the same method  (spectral synthesis) and line list as this work with $\rm{A(Li)}_{\rm{NLTE}}=1.36^{+0.08}_{-0.07}$ dex, which are in reasonable agreement with the results shown in Fig. \ref{li_age}. Notice that \cite{randich/02} observed a star close to the Sun (Cl$^\ast$ NGC 2682 YBP 713, also known as Cl$^\ast$ NGC 2682 SAND 969) and with equivalent width measurements obtained $\rm{A(Li)}_{\rm{NLTE}}=2.06\pm0.10$ dex; however, this star is not a solar twin as it seems more luminous, and also according to \cite{pasquini/08} it is actually a binary system. Thus, this star is disregarded in the following discussion.

Fig. \ref{li_age} shows our sample of M67 solar twins in comparison with the field solar twins sample from \cite{carlos/19}, another solar twin in the M67 open cluster from \cite{liu/16}, and other solar twins that belong to open clusters found in the literature (NGC 2264 from \citealt{king/98}; IC2602 and IC2391 from \citealt{randich/01}; Pleiades from \citealt{soderblom/93}; Blanco 1 from \citealt{ford/05}; NGC 6475 from \citealt{sestito/03}; NGC 1039 from \citealt{jones/97}; Coma Berenices from \citealt{ford/01}; Hyades from \citealt{thorburn/93}; and NGC 762 from \citealt{sestito/04}). There is a recent study of lithium in the cluster Ruprecht 147 \citep{bragaglia/18}; however, all potential solar twins, except one star, have only reported upper limits for lithium abundances.

The advantage of studying an open cluster is the accuracy of stellar age determination, in comparison to field stars age estimates. The age of M67 open cluster adopted here was estimated by the \cite{gaia_col/18}, $\rm{Age}=3.4^{+0.70}_{-0.45}$ Gyr. For comparison, as mentioned in the Introduction, other age estimates from isochrones are around 4 Gyr (e.g. \citealt{castro/11} obtained 3.9$^{+0.6}_{-0.7}$ Gyr), and the age estimated using gyrochronology, i.e. using the rotational period of stars, from the work of \cite{barnes/16} is $4.2 \pm 0.2$ Gyr.


The mean Li abundance value for M67 and its standard deviation is A(Li)$=1.6\pm0.2$ dex, including the solar twin from \cite{liu/16} and excluding the star YBP 1303 due to its upper limit Li abundance. If we consider the four stars, including the upper limit of Li abundance for YBP 1303, we achieve A(Li)$=1.5\pm0.2$ dex. The mean Li abundance found for the M67 open cluster follows the same trend as the field solar twins from \cite{carlos/19}, as expected for solar twins at that age. Interestingly, the scatter estimated from the standard deviation of the Li abundance of the three M67 solar twins (excluding the star with upper limit abundance) is similar to the typical scatter found in the field solar twins sample from \cite{carlos/19}, of about 0.2 dex.

The Li abundance scatter for a given age, as seen in Fig. \ref{li_age}, might be due to their somewhat different characteristics such as rotational periods, metallicities and small difference in masses. These such small differences in their parameters combined can influence in the depth of their convective zones  and their rotational history, thus, resulting in somewhat different amounts of Li burning, as discussed by many studies presented in the literature such as \cite{pinsonneault/90}, \cite{chaboyer/95}, \cite{pace/12} and \cite{somers/17}. Although we work with a sample of solar twins that present very similar characteristics, the Li abundance predicted by the \cite{pace/12} rotational-induced mixing model calibrated to M67 shows that small differences in mass, in the range corresponding to solar twins, imply significant differences in lithium abundances. Thus, part of the scatter could be just a difference in stellar mass.

\section{Conclusions}

We determine Li abundances for three solar twins at the M67 open cluster based on high resolution and high S/N spectra from the GRACES spectrograph at Gemini North.  The mean value of A(Li)$=1.6\pm0.2$ dex at $3.4^{+0.70}_{-0.45}$ Gyr \citep{gaia_col/18} for the M67 solar twins follows the Li-age trend observed by \cite{carlos/16,carlos/19} for field solar twins. 

 

In addition, the scatter found in this work, which is similar to the one found by \cite{carlos/19}, could be caused by small differences of rotational periods, metallicities, and masses on a star-by-star case.

\section*{Acknowledgements}

This study was financed in part by the Coordena\c{c}\~ao de Aperfei\c{c}oamento de Pessoal de N\'ivel Superior - Brasil (CAPES) - Finance Code 001. JM is thankful for the support of Funda\c{c}\~ao de Amparo \`a Pesquisa do Estado de S\~ao Paulo (FAPESP, 2014/18100-4, 2018/04055-8) and Conselho Nacional de Desenvolvimento Cient\'ifico e Tecnol\'ogico (CNPq, Bolsa de Produtividade). JDN and MC acknowledge support from CNPq (Bolsa de Produtividade).

This study is based on observations obtained with ESPaDOnS, located at the Canada-France-Hawaii Telescope (CFHT). CFHT is operated by the National Research Council of Canada, the Institut National des Sciences de l'Univers of the Centre National de la Recherche Scientique of France, and the University of Hawai'i. ESPaDOnS is a collaborative project funded by France (CNRS, MENESR, OMP, LATT), Canada (NSERC), CFHT and ESA. ESPaDOnS was remotely controlled from the Gemini Observatory, which is operated by the Association of Universities for Research in Astronomy, Inc., under a cooperative agreement with the NSF on behalf of the Gemini partnership: the National Science Foundation (USA), the National Research Council (Canada), CONICYT (Chile), Ministerio de Ciencia, Tecnolog\'ia e Innovaci\'on Productiva (Argentina), Minist\'erio da Ci\^encia, Tecnologia e Inova\c{c}\~ao (Brazil), and Korea Astronomy and Space Science Institute (Republic of Korea).




\bibliographystyle{mnras}
\bibliography{mybib} 








\bsp	
\label{lastpage}
\end{document}